\newcommand{\icmltitle}{\mlforastrotitle}
\newcommand{\icmltitlerunning}{\mlforastrotitlerunning} 
\newcommand{\icmlauthor}{\mlforastroauthor}
\newcommand{\icmlaffiliation}{\mlforastroaffiliation} 
\newcommand{\icmlcorrespondingauthor}{\mlforastrocorrespondingauthor}
\newcommand{\icmlkeywords}{\mlforastrokeywords}

\newcommand{\icmlsetsymbol}{\mlforastrosetsymbol}
\newenvironment{icmlauthorlist}
  {\begin{mlforastroauthorlist}}
  {\end{mlforastroauthorlist}}

\documentclass{article}

\usepackage{microtype}
\usepackage{graphicx}
\usepackage{subcaption}
\usepackage{booktabs} 

\usepackage{hyperref}

\usepackage[accepted]{ml4astro2025}

\usepackage{amsmath}
\usepackage{amssymb}
\usepackage{mathtools}
\usepackage{amsthm}

\usepackage[capitalize,noabbrev]{cleveref}

\theoremstyle{plain}

\theoremstyle{definition}

\theoremstyle{remark}

\usepackage[textsize=tiny]{todonotes}

\usepackage{gensymb}
\usepackage{array}
\usepackage{caption}
\usepackage{makecell}
\usepackage{float}
\usepackage{multirow}
\usepackage{threeparttable}
\usepackage{nicefrac}       
\usepackage{xcolor}         

\usepackage{bm}

\def\eqref#1{equation~\ref{#1}}

\def\1{\mathbf{1}}

\def\vd{{\mathbf{d}}}

\def\vn{{\mathbf{n}}}

\def\vx{{\mathbf{x}}}
\def\vy{{\mathbf{y}}}
\def\vz{{\mathbf{z}}}

\DeclareMathAlphabet{\mathsfit}{\encodingdefault}{\sfdefault}{m}{sl}
\SetMathAlphabet{\mathsfit}{bold}{\encodingdefault}{\sfdefault}{bx}{n}

\usepackage{cancel}

\newcommand{\norm}[1]{\left\lVert#1\right\rVert}

\icmltitlerunning{A Fast Generative Framework for High-dimensional Posterior Sampling}

\begin{document}

\twocolumn[
\icmltitle{A Fast Generative Framework for High-dimensional Posterior Sampling: \\ Application to CMB Delensing}



\icmlsetsymbol{equal}{*}

\begin{icmlauthorlist}
\icmlauthor{Hadi Sotoudeh}{udem,mila,ciela}
\icmlauthor{Pablo Lemos}{udem,mila,ciela}
\icmlauthor{Laurence Perreault-Levasseur}{udem,mila,ciela,cca}
\end{icmlauthorlist}

\vspace{0.1em}
\begin{center}
Simons Collaboration on \textit{Learning the Universe}
\end{center}
\vspace{-1.1em}

\icmlaffiliation{udem}{Department of Physics, Université de Montréal, Montréal, Canada}
\icmlaffiliation{mila}{Mila - Quebec AI Institute, Montréal, Canada}
\icmlaffiliation{ciela}{Ciela, Montreal Institute for Astrophysics and Machine Learning, Montréal, Canada}
\icmlaffiliation{cca}{Center for Computational Astrophysics, Flatiron Institute, New York, USA}

\icmlcorrespondingauthor{Hadi Sotoudeh}{hadi.sotoudeh@umontreal.ca}

\icmlkeywords{High-dimensional Bayesian Inference, Deep Generative Models, Uncertainty Quantification, Cosmology, CMB}

\vskip 0.6in
]



\printAffiliationsAndNotice{}  

\begin{abstract}

We introduce a deep generative framework for high-dimensional Bayesian inference that enables efficient posterior sampling. As telescopes and simulations rapidly expand the volume and resolution of astrophysical data, fast simulation-based inference methods are increasingly needed to extract scientific insights. While diffusion-based approaches offer high-quality generative capabilities, they are hindered by slow sampling speeds. Our method performs posterior sampling an order of magnitude faster than a diffusion baseline. Applied to the problem of CMB delensing, it successfully recovers the unlensed CMB power spectrum from simulated observations. The model also remains robust to shifts in cosmological parameters, demonstrating its potential for out-of-distribution generalization and application to observational cosmological data.

\end{abstract}


\vspace{-\baselineskip}
\section{Introduction}

The volume and complexity of astrophysical data are rising significantly due to modern instruments and simulations \citep{bellm2019_scheduling, ntampaka2021_ml}. Space missions and ground-based telescopes such as the James Webb Space Telescope \citep{gardner2006_jwst}, Euclid \citep{racca2016_euclid}, Square Kilometre Array \citep{ska}, LSST \citep{lsst}, Simons Observatory \citep{galitzki2024_simons}, and CMB-S4 \citep{abazajian2016_cmbs4} will not only bring about exponential growth to the size of the observed data but will also increase the resolution and data acquisition rate to unprecedented levels.

Bayesian inference provides a principled framework for estimating physical parameters from observations. However, the scale and dimensionality of modern datasets pose substantial challenges for modeling the posterior distribution. For likelihood-based models, inference can become intractable because the likelihood function often breaks down as its simplifying assumptions fail at higher data fidelity. Even moving to likelihood-free (or implicit-likelihood) approaches, which bypass the need for an explicit likelihood by relying on forward simulations, fails to fully address key bottlenecks. Specifically, the computational cost of posterior sampling often grows prohibitively with data complexity.


In recent years, machine learning (ML) methods have become increasingly popular for addressing challenges in Bayesian inference. These techniques have driven significant advancements in simulation-based inference (SBI) approaches \citep{cranmer2020_sbi, durkan2020_contrastivelfi}. However, several challenges remain in realizing the full potential of these methods for scientific discovery. Key obstacles include obtaining reliable uncertainty estimates, generalizing to real data, developing training schemes that work across problems with minimal fine-tuning, and enabling fast inference.

This paper introduces a deep generative framework for fast, high-dimensional Bayesian inference. Our approach builds on the Hierarchical Probabilistic U-Net (HPU-Net) architecture \citep{kohl2019_hierarchical}, which enables learning structured probabilistic representations using the variational autoencoder (VAE) inference framework \citep{kingma2022_vae}. The VAE-based design allows for fast sampling. We demonstrate the framework's practical utility on the task of CMB delensing — removing the effects of weak gravitational lensing from the observed cosmic microwave background — and show that the model rapidly and reliably recovers the primordial\footnote{Here, \textit{primordial} refers to the CMB signal in the absence of gravitational lensing.} CMB power spectrum with reasonable uncertainty estimates.

\clearpage



\section{Methods}

\subsection{Architecture}

Our inference framework draws samples from the high-dimensional posterior distribution $p(\vx | \vy)$ of parameters $\vx$ given the observed data $\vy$.

To make the problem more tractable, we separate the task of learning the posterior mean from modeling the dispersion of samples around the mean. This decomposition enables using lighter-weight probabilistic networks, which are typically more expensive to train than deterministic ones. This is achieved through two seperately trained networks:

\begin{enumerate}
    \item The \textit{Mean Network} learns the posterior mean,\\
    $ \bar{\vx}(\vy) = \mathbb{E}_\vx \! \Big[ \, p(\vx | \vy) \, \Big]$.

    \item The \textit{Dispersion Network} models the variability of posterior samples around the mean. It generates samples from the distribution of deviations, defined as the difference between a posterior sample and the mean, $\bm\delta := \vx - \bar{\vx}$.  
\end{enumerate}

The Mean Network follows the U-Net architecture \citep{ronneberger2015_unet} and produces deterministic outputs. The Dispersion Network can be any generative model; we use the Hierarchical Probabilistic U-Net (HPU-Net) \citep{kohl2019_hierarchical}, which introduces stochasticity by adding sampling layers to the expanding path of the U-Net.  

\subsection{Training Objective}

\subsubsection*{Mean Network}

The Mean Network minimizes the Mean Squared Error (MSE) between its output $\hat{\bar{\vx}}$ and the target $\vx$:
\begin{equation}
    \ell_{\text{MeanNet}} = \frac{1}{2}
    \text{ } \norm{\hat{\bar{\vx}} - \vx}_2^2 \; .
\end{equation}
According to \citet{adler2018_msemean}, the optimal solution of training a deterministic neural network (i.e., one without sampling layers) using MSE loss is the posterior mean. 
Hence, if properly trained, the Mean Network is guaranteed to learn the posterior mean.

\subsubsection*{Dispersion Network}

The Dispersion Network models the joint distribution of deviations $\bm\delta$ and some latent variables $\vz$, conditioned on observed data $\vy$: $ p(\bm\delta, \vz | \vy) = p(\vz | \vy) \; p(\bm\delta | \vz, \vy) $. It is trained similarly to a VAE, where the network is tasked with maximizing the evidence lower bound (ELBO), i.e., minimizing 
\begin{equation} \label{eqn:dipersionnet_loss}
    \begin{split}
        \ell_\text{DispersionNet}
        &=
        \mathbb{E}_{\vz \sim q(\vz | \bm\delta,\vy)}
        \Bigl[ - \ln p(\bm\delta | \vz,\vy) \Bigr]
        \\
        &\qquad \;\;\, + \;
        D_\text{KL} \Bigl[ \, q(\vz | \bm\delta,\vy) \parallel p(\vz | \vy) \, \Bigr]
        \\
        &=
        \ell_\text{rec} + \ell_\text{KL} \; .
    \end{split}
\end{equation}
The first term is the reconstruction loss, which measures how well the generated data matches the expected output. The second term is the Kullback–Leibler (KL) divergence between the variational posterior $q(\vz | \bm\delta,\vy)$ and the prior $p(\vz | \vy)$. For further details, see \citet{kohl2019_hierarchical}.

We employ a diagonal Gaussian negative log-likelihood reconstruction loss, 
\begin{equation} \label{eqn:dipersionnet_gnll}
    \begin{split}
        \ell_\text{rec}
        &= \frac{1}{2} \bigl( \ln|\hat{\bm\Sigma}|
            + (\bm\delta - \hat{\bm\mu})^\top \hat{\bm\Sigma}^{-1} (\bm\delta - \hat{\bm\mu})
          \bigr)
        \\
        &=
        \frac{1}{2} \sum_i{
        \Bigl[
            \ln ( \hat\sigma^2_i ) +
            \frac{ ( \delta_i - \hat{\mu}_i )^2 }{ \hat\sigma_i^2 }
        \Bigr]
        } \; ,
    \end{split}
\end{equation}
where $\hat{\bm\mu}$ and $\hat{\bm\Sigma}$ (or $\hat{\mu}_i$ and $\hat\sigma^2_i$) are \textbf{estimated from multiple samples generated by the Dispersion Network}, and $\bm\delta = \vx - \hat{\bar{\vx}}$ is the deviation of the ground-truth target from MeanNet's output. The sum is over all output pixels $i$.  

\begin{figure*}[h]
    \centering
    
    \begin{subfigure}{0.61\textwidth}  
        \centering
        \includegraphics[width=\textwidth]{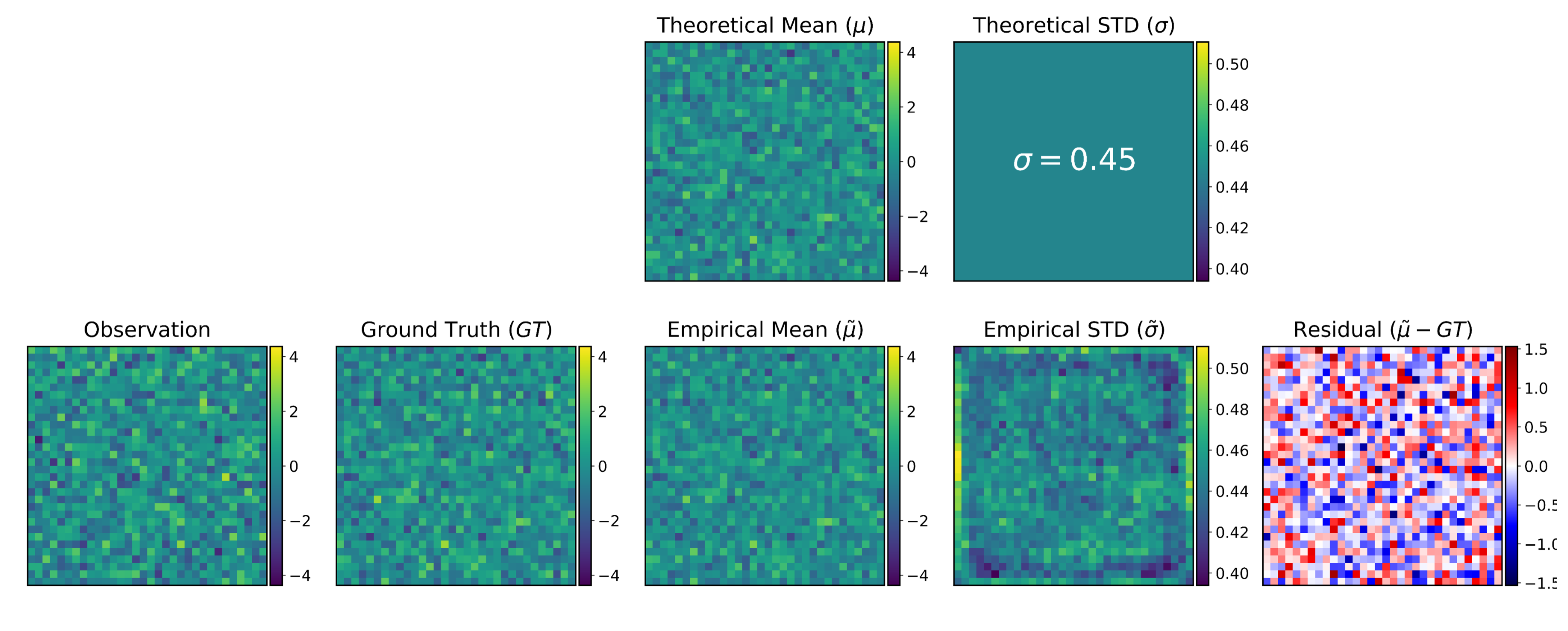}
        \caption{}
        \label{fig:grf_results_moment}
    \end{subfigure}
    \hspace{0.03\textwidth}
    \begin{subfigure}{0.265\textwidth}  
        \centering
        \raisebox{-0.6\baselineskip}[0pt][0pt]{ 
            \includegraphics[width=\textwidth]{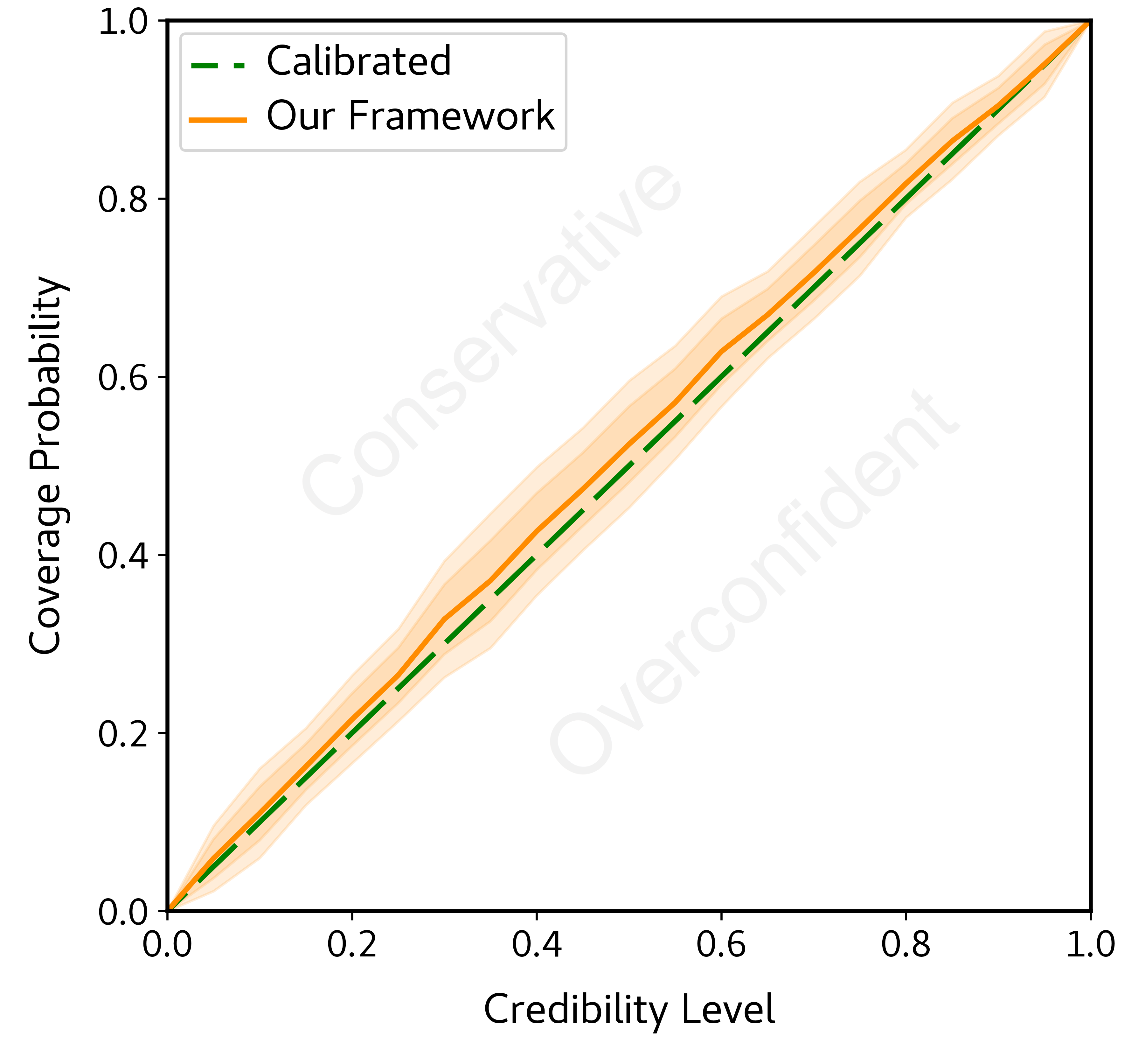}
        }
        \caption{}
        \label{fig:grf_results_coverage}
    \end{subfigure}
    \vspace{-0.3\baselineskip}
    
    \caption{GRF rotation results:  (a) Comparing empirical and theoretical moments for a test example. The pixel-wise empirical means and standard deviations are estimated using 1000 posterior samples.  (b) TARP coverage test results.}
    \label{fig:grf_results}

\end{figure*}

This choice of reconstruction loss offers two advantages:
(i) For networks with \textit{leaky paths}\footnote{A \textit{leaky path} is a route in the network's computation graph that allows information to flow from the ground truth to the output without passing through the latent space - for example, skip connections in the U-Net architecture.}, maximizing the ELBO can lead to the KL vanishing problem \citep{bowman2016_klvanish, fu2019_cyclicalannealing}, causing the network to ignore the latent variables and collapse to a deterministic solution. Maximizing the ELBO under the proposed reconstruction loss does not suffer from this problem; (ii) In VAE-like frameworks, \textit{directly optimizing} the variance $\hat{\sigma}_i^2$ can cause it to collapse to zero \citep{rezende2018_taming}, resulting in a deterministic model. By estimating the variance from posterior samples, we avoid this collapse while including a loss term that provides a useful signal for calibrating the model's uncertainty estimates. A key component of this design is the VAE-based approach, which allows fast sampling and makes training feasible. In contrast, implementing this strategy with a diffusion model would be significantly more expensive.

\subsection{Baseline}

For some experiments, we compare performance against a baseline in which the Dispersion Network is replaced with a conditional diffusion model, trained using the denoising diffusion probabilistic model (DDPM) framework of \citet{ho2020_diffusion}.

Further details on the architecture, training procedure, and baseline can be found in Appendix~\ref{app:b}.


\section{Experiments and Results}

\subsection{ \scalebox{0.93}[1]{Problem 1: Rotating Gaussian Random Fields (GRFs)} }

This experiment evaluates our framework on a problem with an analytically tractable posterior: a linear inverse problem, where the data $\vy$ and parameters $\vx$ are related through $ \vy = \mathbf{R} \vx + \vn $. Here, $\mathbf{R}$ is a transformation matrix, and $\vn$ is a random noise vector. If the parameter prior $p(\vx)$ and the noise distribution $p(\vn)$ are both Gaussian, then the posterior distribution $p(\vx | \vy)$ is also Gaussian \citep{kaipio_invgaussian}, with the following parameters:
\begin{equation}
    \begin{split}
        \bm\mu_\text{post}
        =
        (\mathbf{M} + \mathbf{M}^\top)^{-1} \, \vd
        \qquad \qquad
        \bm\Sigma_\text{post}^{-1}
        =
        \mathbf{M},
        \\
        \scalebox{0.7}{$
        \vd := 2 \; ( \bm\Sigma_\text{pri}^{-1} \: \bm\mu_\text{pri}   +   \mathbf{R}^\top \bm\Sigma_\text{n}^{-1} \vy )
        \qquad \quad \;\;
        \mathbf{M} := ( \bm\Sigma_\text{pri}^{-1}   +   \mathbf{R}^\top \bm\Sigma_\text{n}^{-1} \mathbf{R} )
        $}
    \end{split}
\end{equation}
where the \textit{pri}, \textit{post}, and \textit{n} subscripts indicate the prior, posterior, and noise distributions, respectively. We set $\mathbf{R}$ to be a $90\degree$ rotation. To evaluate performance on this task, we compare $\bm\mu_\text{post}$ and $\bm\Sigma_\text{post}$ with the empirical mean and covariance estimated from the model's predictions (Figure~\ref{fig:grf_results_moment}). We also employ the Test of Accuracy with Random Points (TARP) \citep{lemos2023_tarp} to assess the calibration of the credible regions produced using the model (Figure~\ref{fig:grf_results_coverage}).

\subsection{Problem 2: CMB Delensing}

CMB lensing occurs when the cosmic microwave background is distorted by weak gravitational lensing from intervening cosmic structures between the last scattering surface and Earth \citep{lewis2006_cmblensing, hanson2010_cmblensing}. Correcting for this effect is crucial for obtaining unbiased estimates of cosmological parameters, studying the distribution of matter in the universe, and removing the lensing-induced B-mode pattern, which can be confused with signals from primordial gravitational waves. Recent CMB lensing results and analyses are presented in \citet{plancklensing2020, lensing_spt, lensing_act}. Lately, deep learning has also been applied to CMB delensing. Specifically, \citet{caldeira2019_deepcmb, li2022_delensing, yan2023_mimounet, yan2023_rdlfunet} used convolutional neural networks to produce point estimates of unlensed CMB maps. More recently, \citet{floss2024diffusiondelensing} proposed a diffusion-based approach.  

\begin{figure*}[h]
    \centering
    \begin{subfigure}{0.605\textwidth}  
        \centering
        \raisebox{-0.5\baselineskip}[0pt][0pt]{ 
            \includegraphics[width=\textwidth]{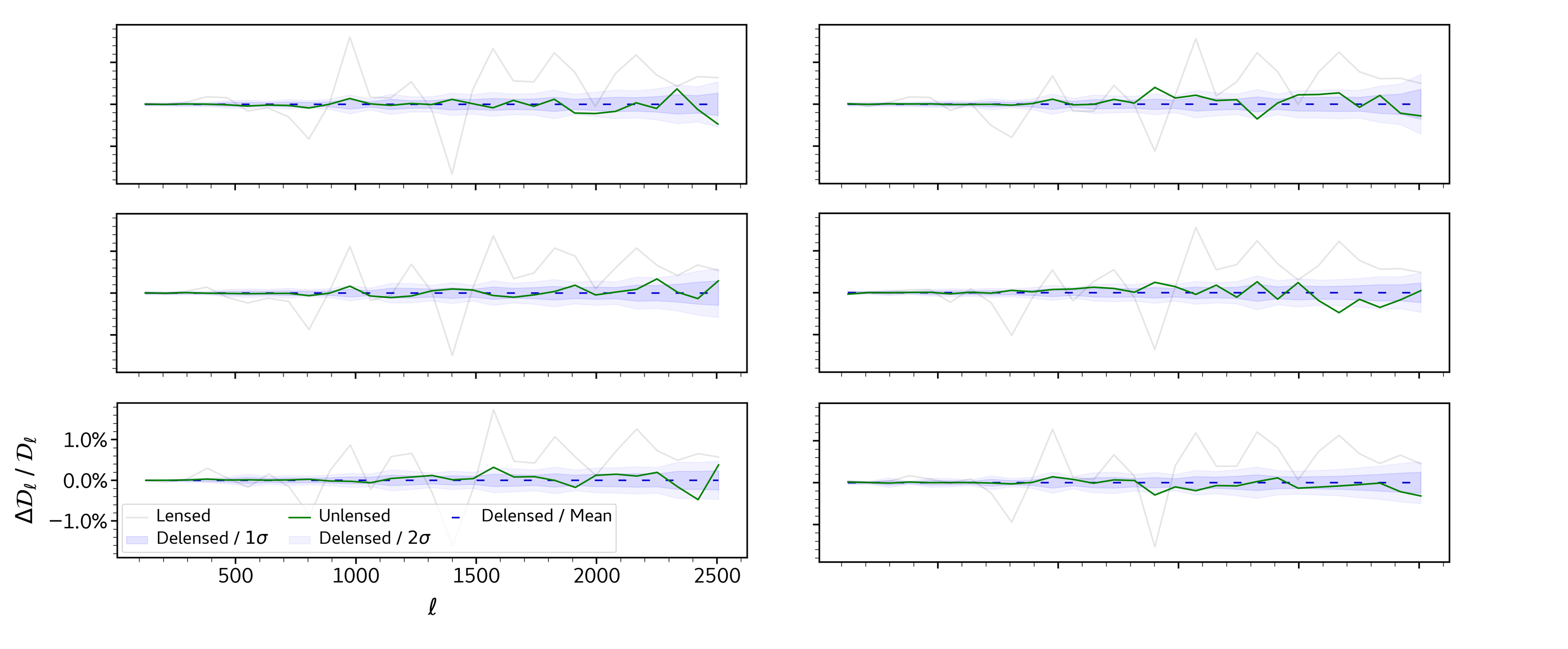}
        }
        \caption{}
        \label{fig:cmb_results_ps}
    \end{subfigure}
    \hspace{0.03\textwidth}
    \begin{subfigure}{0.265\textwidth}  
        \centering
        \includegraphics[width=\textwidth]{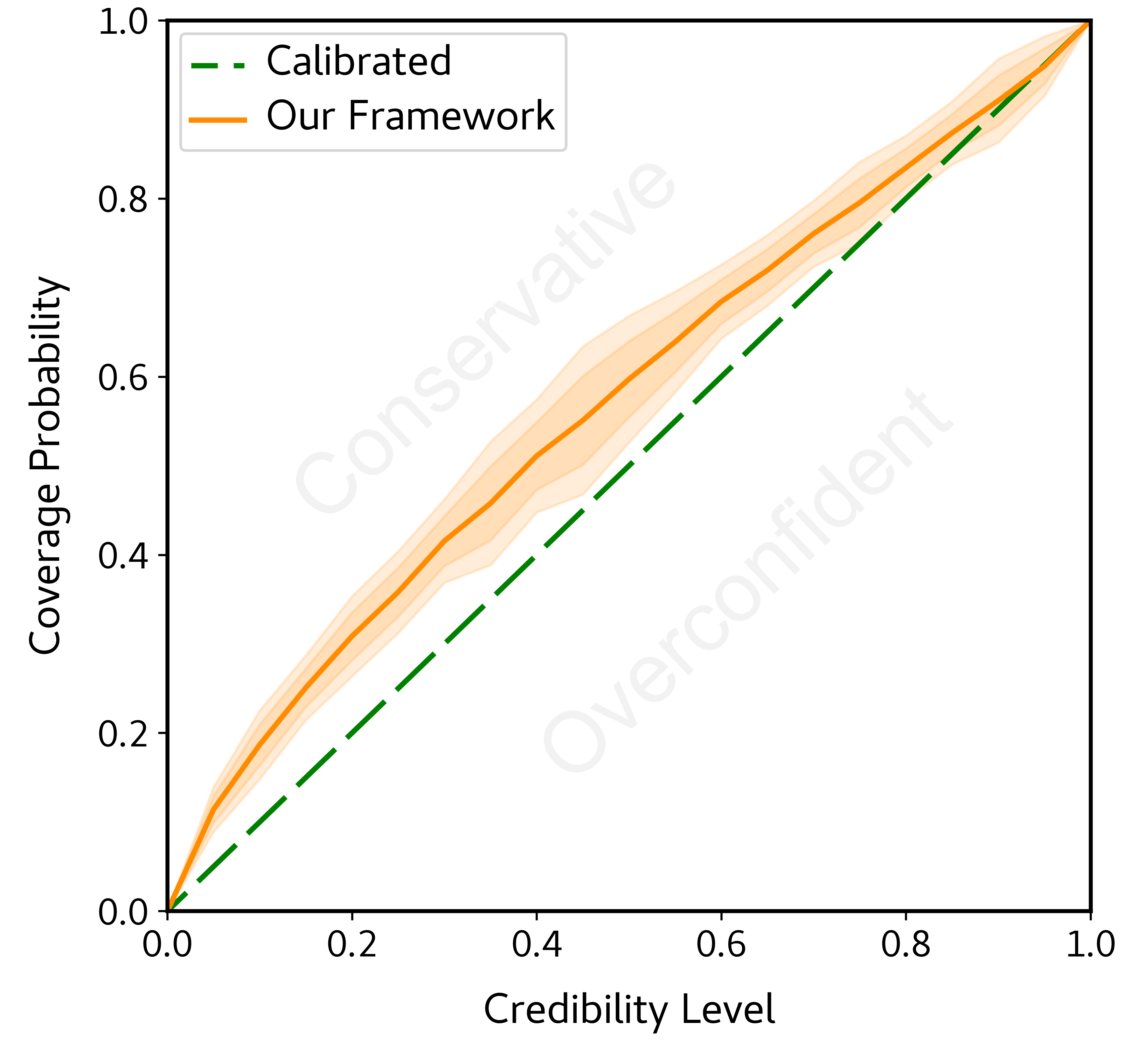}
        \vspace{-1.5\baselineskip}
        \caption{}
        \label{fig:cmb_results_coverage}
    \end{subfigure}

    \vspace{0.5\baselineskip}

    \begin{subfigure}{\textwidth}
        \centering
        \includegraphics[width=0.91\textwidth]{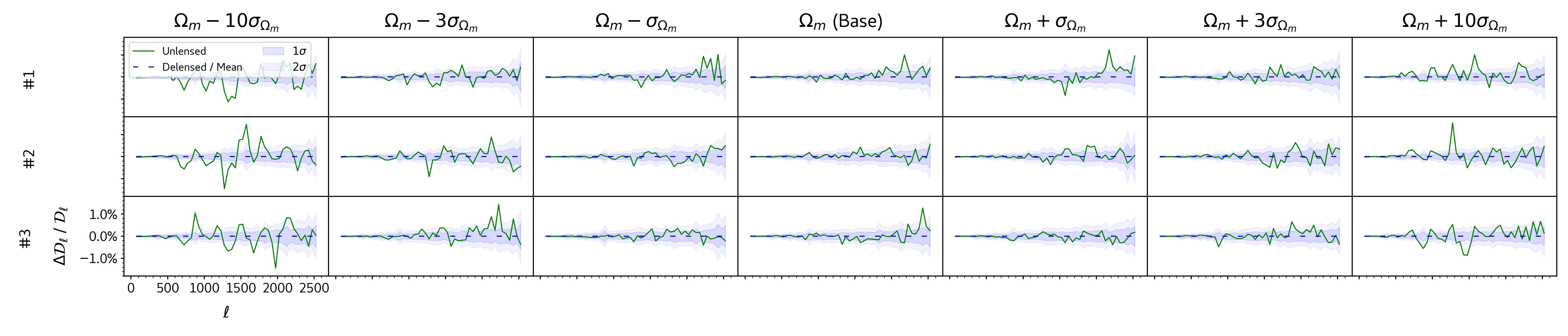}
        \vspace{-1.5\baselineskip}
        \caption{}
        \label{fig:cmb_results_ood}
    \end{subfigure}

    \vspace{-0.3\baselineskip}
    
    \caption{CMB delensing results:  (a) Comparison of the lensed (input), unlensed (target), and delensed (obtained from model outputs) TT power spectra for six test examples. The delensed spectrum is represented by mean and uncertainty regions computed from 1000 posterior samples. Relative differences $\Delta\mathcal{D}_\ell / \mathcal{D}_\ell$ are computed with respect to the mean delensed spectrum.  (b) TARP coverage test results.  (c) Out-of-distribution performance for different $\Omega_m$ values, with all other cosmological parameters fixed. Each column corresponds to an $\Omega_m$ value, which differs from the training (base) value by a factor of the Planck satellite measurement error $\sigma_{\Omega_m}$. Each row corresponds to a unique noise realization used to generate the lensed and unlensed maps. }
    \label{fig:cmb_results}
    \vspace{-0.5\baselineskip}

\end{figure*}

In our delensing experiments, we use the Code for Anisotropies in the Microwave Background (CAMB) \citep{lewis2002_camb} to generate simulated CMB observations, with and without lensing effects. Specifically, we use the lensed and unlensed TT power spectra from CAMB to generate realizations of CMB maps corresponding to each spectrum. The inference framework is tasked with predicting the difference between the lensed and unlensed maps, given the lensed map as input. To evaluate performance on this task, we compare the power spectrum of the delensed CMB maps with the target (unlensed) power spectrum (Figure~\ref{fig:cmb_results_ps}). We also use TARP to assess the model's uncertainty calibration (Figure~\ref{fig:cmb_results_coverage}). Finally, to test robustness, we evaluate the model on out-of-distribution (OOD) CMB maps generated using different cosmological parameters (Figure~\ref{fig:cmb_results_ood}). To construct an OOD dataset, we retain the same noise realizations as in the in-distribution dataset; however, we vary the matter density parameter $\Omega_m$ by a multiple of the Planck satellite measurement uncertainty $\sigma_{\Omega_m} = 0.0073$, which affects the TT power spectra generated by CAMB. Using this procedure, we generate several OOD datasets with different $\Omega_m$ values.

Further details on the data generation procedure can be found in Appendix~\ref{app:a}.

\section{Discussion and Conclusion}

Our framework closely approximates the posterior distribution, with slightly conservative credible intervals. In the GRF rotation experiment (Figure~\ref{fig:grf_results}), the empirical and theoretical moments align closely, and the model's uncertainty estimates are well calibrated. For CMB delensing (Figure~\ref{fig:cmb_results_ps}), the unlensed spectrum consistently falls within the uncertainty regions computed from the power spectrum of posterior samples. Furthermore, the model generalizes well to out-of-distribution (OOD) inputs (Figure~\ref{fig:cmb_results_ood}): unless the matter density parameter $\Omega_m$ is significantly perturbed from training values, the predicted uncertainty bands continue to contain the true unlensed spectrum. This indicates potential for application to other OOD scenarios, most importantly for generalizing from simulations to observations.  

\vspace{-\baselineskip}
\begin{table}[H]
	\centering
    \caption{Comparison of the time required to generate 50 posterior samples for a batch of 4 test examples on an NVIDIA H100 GPU. Uncertainties are estimated from 10 runs. The numbers next to the diffusion baselines indicate the number of denoising steps.}
    \vspace{0.5\baselineskip}
	\setcellgapes{4pt}
	\makegapedcells
    \footnotesize
	\begin{tabular}{ r | c }
		  \textbf{Model}  & \textbf{Sampling Time (s)}  \\  \Xhline{1pt}
		  Our Framework     &     $0.31 \pm 0.05$             \\  \cline{1-2}
        Diffusion Baseline-100        &     $12.85 \pm 0.09$ \;             \\ \cline{1-2}
        Diffusion Baseline-1000        &     $125.6 \pm 0.4$ \;\;\,             \\
	\end{tabular}
	\label{tbl:sampling_time}
\end{table}
\vspace{-\baselineskip}


Furthermore, our inference framework provides uncertainty estimates, offering an advantage over deep learning approaches to CMB delensing that produce only point estimates. In addition, leveraging a VAE-based architecture enables fast inference \citep{vae_faster}. As shown in Table~\ref{tbl:sampling_time}, our model is at least 40 times faster than diffusion-based baselines. Beyond their slow sampling, conditional diffusion models are also known to produce overconfident uncertainty estimates \citep{tachella2023_overconfident, bourdin2024_inpainting}. By comparison, our model's uncertainty estimates are slightly conservative. Lastly, the proposed reconstruction loss prevents KL vanishing and enables implicit uncertainty calibration, ensuring latent variables remain informative throughout training and preventing collapse to a deterministic solution.



Future directions include improving uncertainty calibration, scaling the framework to higher-dimensional problems, applying it to observational data, and conducting more extensive comparisons against baselines. To improve calibration, one could explore modified loss functions that encourage predicting extreme values, thereby reducing the conservativeness of credible intervals. Scaling to higher dimensions would better highlight the advantages of this approach over diffusion models, particularly as the sampling speed gap becomes more pronounced. Applying the framework to observational data includes both training directly on real observations and investigating how models trained on simulations generalize — or transfer — to real-world settings.

In this paper, we presented a framework for high-dimensional posterior sampling based on the Hierarchical Probabilistic U-Net. This method is fast, robust, and grounded in a sound theoretical foundation. It is well-suited for simulation-based inference scenarios where direct likelihood modeling is infeasible, enabling the characterization of high-dimensional posteriors and uncertainty estimation in physical measurements using neural networks.



\section*{Acknowledgements}

This work was conducted as part of the Simons Collaboration on \textit{Learning the Universe}.
It was enabled by computational resources provided by Calcul Québec, Compute~Canada, and the Digital Research Alliance of Canada.

Hadi~Sotoudeh acknowledges support from the IVADO MSc Excellence Scholarship and gratefully thanks the Institute of Astronomy at the University of Cambridge, Girton College, and Steven Dillmann for their support in the preparation of this work for the ML4Astro workshop. We thank the reviewers for their valuable comments.

\section*{Impact Statement}

This paper presents work whose goal is to advance the field of 
Machine Learning and its applications in Astrophysics. There are many potential societal consequences 
of our work, none which we feel must be specifically highlighted here.


\bibliography{biblio}
\bibliographystyle{icml2025}

\newpage
\appendix
\onecolumn


\section{Data Generation} \label{app:a}

\vspace{\baselineskip}
\subsection{GRF Rotation}

\begin{figure}[h]
    \centering
    \includegraphics[width=0.65\textwidth]{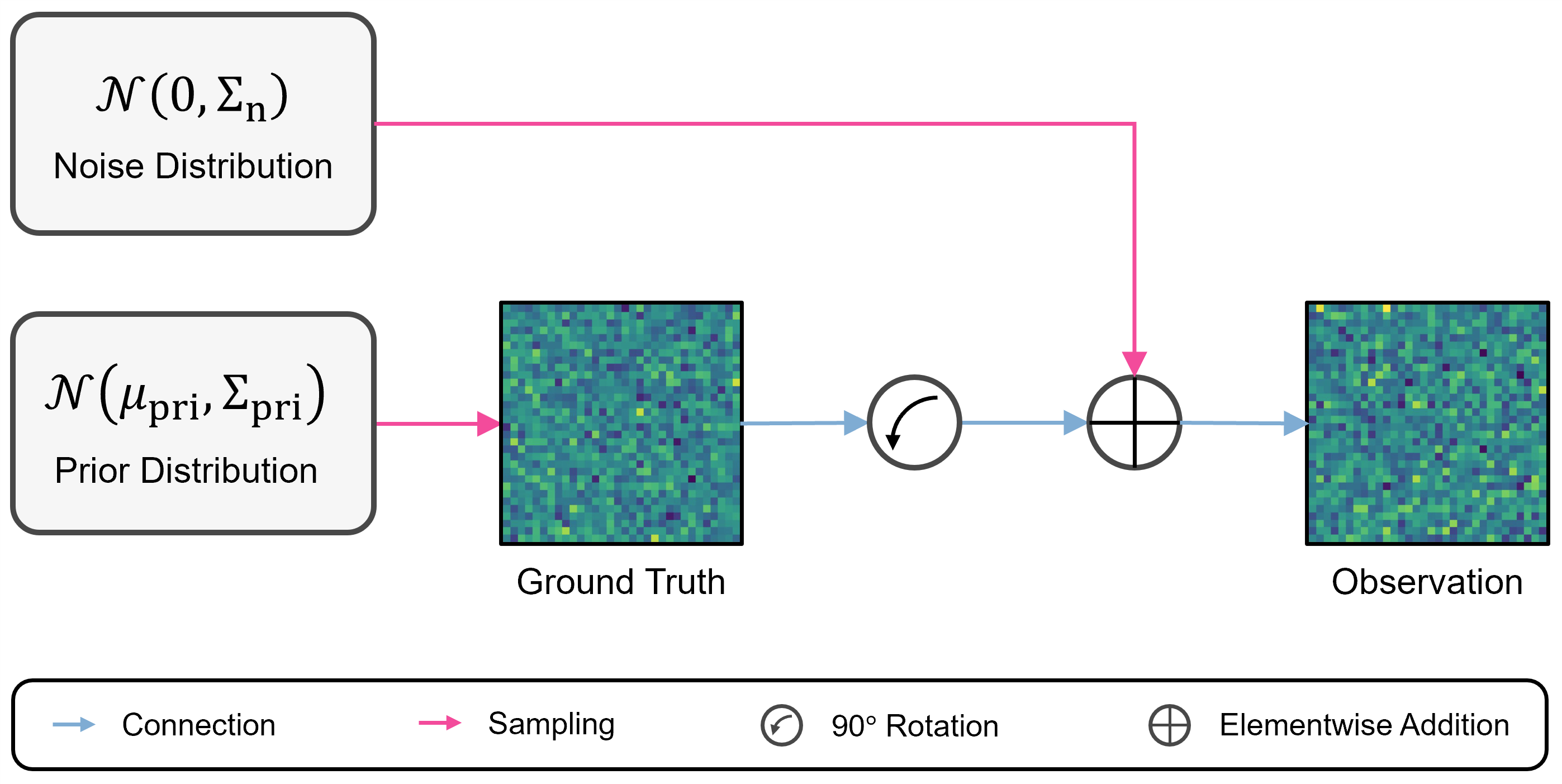}
    \vspace{0.3\baselineskip}
    \caption{Data generation process for the GRF rotation experiment.}
    \label{fig:grf_data}
\end{figure}

Figure \ref{fig:grf_data} illustrates the data generation process for the GRF rotation experiment. We train (test) the model on $2^{16}$ ($2^{11}$) examples with $32 \times 32$ resolution. The data was generated by sampling from the prior distribution ($\mu_\text{pri}=0, \sigma_\text{pri}=1$), rotating the samples by $90^\circ$, and adding noise ($\sigma_\text{n}=0.5$) to the rotated samples. In order to increase the effective size of the training set and enhance the model's generalization, each training epoch uses different noise realizations.

\subsection{CMB Delensing}

\begin{figure}[h]
    \centering
    \includegraphics[width=0.98\textwidth]{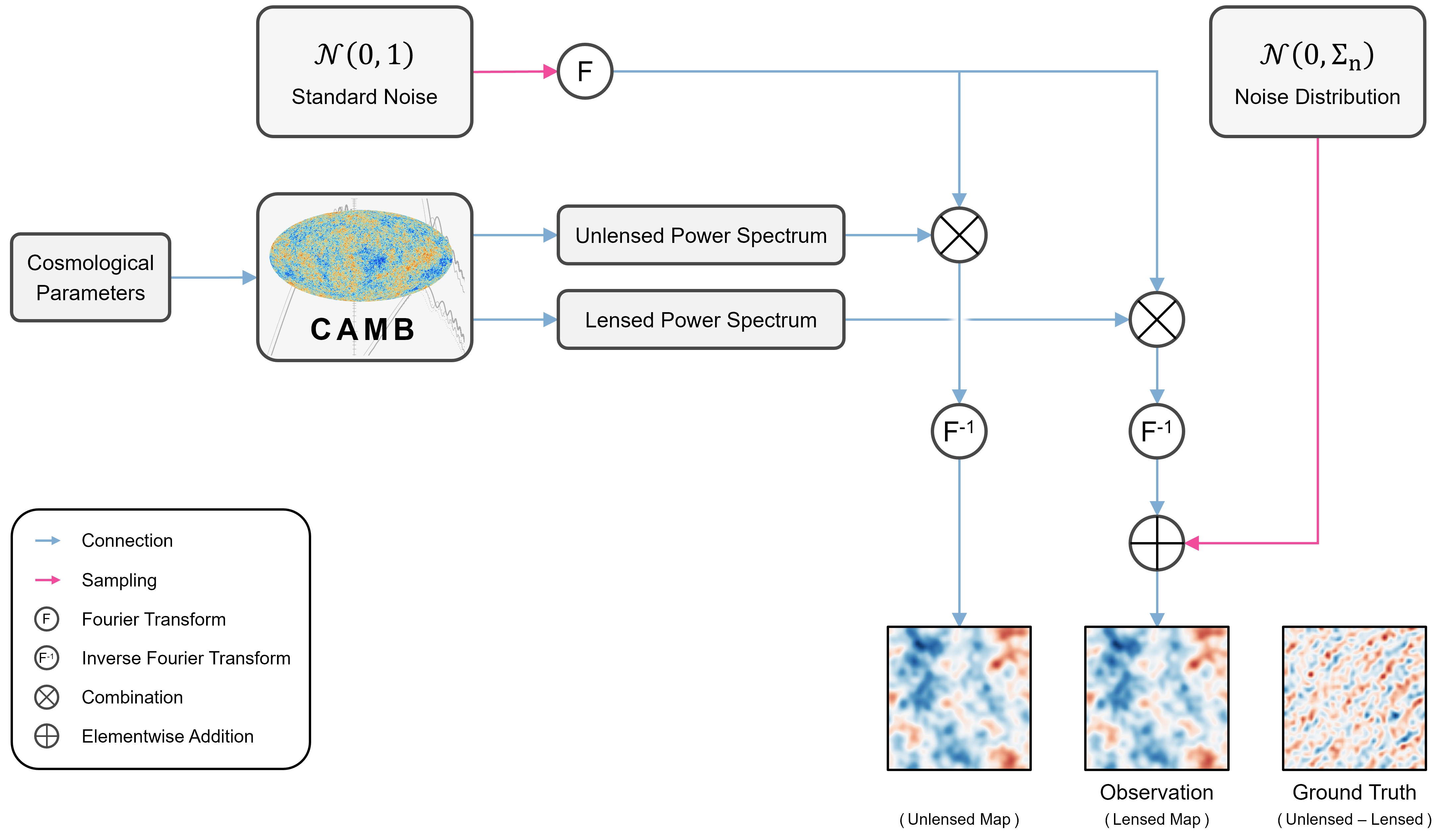}
    \vspace{-0.3\baselineskip}
    \caption{Data generation process for the CMB delensing experiment.}
    \label{fig:cmb_data}
\end{figure}

Figure \ref{fig:cmb_data} illustrates the data generation process for the CMB delensing experiment. We begin by calculating the lensed and unlensed TT power spectra using CAMB, based on the cosmological parameters listed in Table \ref{tbl:cosmo_params}. We then apply the same noise realization to these power spectra to generate lensed and unlensed maps in Fourier space, and then transform the resulting maps to real space. The training (test) set contains $2^{13}$ ($2^{11}$) maps, each covering a $160^\prime \times 160^\prime$ region with $32 \times 32$ resolution. To simulate observational noise, we add random noise ($\sigma_\text{n}=20$) to the lensed maps. Similar to the GRF rotation experiment, we apply noise during training and use different realizations for each epoch. The input to the framework is a lensed map, and the target is the \textbf{difference} between the lensed and unlensed maps.

\begin{table}[H]
	\centering
    \vspace{0.5\baselineskip}
    \caption{Cosmological parameters for the simulated CMB maps.}
    \vspace{\baselineskip}
	\setcellgapes{6pt}
	\makegapedcells
    \small
	\begin{tabular}{ r l | c }
		  \multicolumn{2}{c | }{\textbf{Parameter}}  & \textbf{Value}  \\   \Xhline{1pt}
		$H_0$ &   Hubble parameter ($\text{km } \text{s}^{-1} \text{Mpc}^{-1}$)               &     67.5 \\ \cline{1-3}
        $\Omega_m$  &   Matter density parameter      & 0.27\\ \cline{1-3}
        $\Omega_k$  &   Curvature  density parameter          & 0  \\ \cline{1-3}
        $\tau$  &   Thomson scattering optical depth              & 0.06  \\ \cline{1-3}
        $\ln{ ( 10^{10} A_s ) }$    &   Primordial curvature perturbations  & 2.996 \\ \cline{1-3}
        $n_s$   &   Scalar spectrum power-law index               & 0.965  \\ \cline{1-3}
        $\Sigma m_\nu$  &   Sum of active neutrino masses (eV)      & 0.06  \\ \cline{1-3}
        $r$ &   Tensor-to-scalar ratio          & 0  \\
	\end{tabular}
	\label{tbl:cosmo_params}
\end{table}

\section{Training} \label{app:b}

\subsection{Hyperparameters}

The following hyperparameters were used for the experiments presented in the results:

\textbf{Data Normalization.} We use Gaussian standardization for the GRF rotation experiment and maximum absolute scaling for the CMB delensing experiment. The normalization coefficients are calculated globally, i.e., based on the entire dataset rather than individually for each example. We use the training set's normalization coefficients during inference.

\textbf{Architecture.} All U-Nets have a depth of 4, with a downsampling and upsampling scale factor of 2. For $32 \times 32$ inputs, this results in a feature map dimensionality of $2 \times 2$ at the U-Net bottleneck. We employ average pooling for downsampling and nearest-neighbor interpolation for upsampling, except for the GRF rotation Mean Network, which uses strided convolution and transpose convolution. The number of channels for different models is summarized in Table \ref{tbl:architecture}. All models use the following kernel sizes for their respective convolutional layers: [7, 7, 7, 5, 3].

\vspace{-0.5\baselineskip}
\begin{table}[H]
	\centering
    \begin{threeparttable}
    \caption{Model architecture.}
    \vspace{\baselineskip}
	\setcellgapes{6pt}
	\makegapedcells
    \small
	\begin{tabular}{ r l | c | l }
		  \multicolumn{2}{c | }{\textbf{Model}}  & \textbf{Input Channels}  &  \textbf{Intermediate Channels$^+$}    \\   \Xhline{1pt}
		\multirow{2}{*}{GRF Rotation} &   Mean Network               &     1  &  [32, 64, 128, 128, 128]   \\ \cline{2-4}
          &   Dispersion Network      & 2$^*$  &  [8, 16, 32, 32, 32]      \\ \Xhline{1pt}
        \multirow{2}{*}{CMB Delensing}  &   Mean Network          & 1  &  [64, 128, 256, 256, 256]    \\ \cline{2-4}
         &   Dispersion Network          & 2$^*$  &  [16, 32, 64, 64, 128]    \\
	\end{tabular}
    \begin{tablenotes}
        \item[+] Ordered from input to bottleneck. The expanding path mirrors this structure.
        \item[*] The Dispersion Network receives both the observation and the posterior mean.
    \end{tablenotes}
    \label{tbl:architecture}
	\end{threeparttable}
\end{table}
\vspace{-0.5\baselineskip}

\textbf{Loss Function.} When training the Dispersion Networks, we draw 10 samples from the model to estimate the posterior mean and covariance.

\textbf{Optimizer.} We employ the Adamax optimizer with a weight decay of $10^{-5}$ and an initial learning rate of $10^{-4}$. All other optimizer hyperparameters are set to their PyTorch default values. The learning rate is then reduced by a factor of 5 at specific milestones: epochs 200 and 300 for the Mean Networks, every 20 epochs for the GRF rotation Dispersion Network, and every 250 epochs for the CMB delensing Dispersion Network. We use a batch size of 128.

\subsection{Compute Resources}

Table \ref{tbl:training_time} summarizes the model size and training time on a single NVIDIA V100 GPU. The training set size and number of epochs are also included to ensure meaningful comparisons.

\begin{table}[H]
	\centering
    \caption{Model size and training time.}
    \vspace{\baselineskip}
	\setcellgapes{6pt}
	\makegapedcells
    \small
	\begin{tabular}{ r l | c | c | c | c }
		  \multicolumn{2}{c | }{\textbf{Model}}  & \textbf{\# Parameters}  &  \textbf{Dataset Size}  &   \textbf{Epochs}  &   \textbf{Training Time}  \\   \Xhline{1pt}
		\multirow{2}{*}{GRF Rotation} &   Mean Network               &     7.5 M  &  $2^{16}$  &  400  &     25 hours \\ \cline{2-6}
          &   Dispersion Network      & 780 k  &  $2^{16}$  &  50   & 13 hours \\ \Xhline{1pt}
        \multirow{2}{*}{CMB Delensing}  &   Mean Network          & 27.8 M  &  $2^{13}$  &  400  &   87 hours  \\ \cline{2-6}
         &   Dispersion Network          & 3.9 M  &  $2^{13}$  &  600  &   42 hours  \\
	\end{tabular}
	\label{tbl:training_time}
\end{table}

\subsection{Baseline}

We establish our diffusion baselines using the HuggingFace Diffusers library \citep{diffusers}, specifically instantiating the \texttt{UNet2DConditionModel}. This architecture follows a symmetrical encoder-decoder U-Net design. The encoder consists of three standard convolutional blocks, followed by a cross-attention block that integrates conditioning information. The decoder mirrors this structure in reverse. The number of channels increases progressively through the convolutional blocks: 64, 128, and 256. The conditioning inputs, comprising the observation $\vy$ and the Mean Network's prediction $\hat{\bar{\vx}}$, are processed by a separate convolutional encoder. This encoder is composed of three sequential convolutional blocks. The resulting feature maps are restructured and used as keys and values in the U-Net’s cross-attention mechanism.

We train this model using the denoising diffusion probabilistic model (DDPM) framework of \citet{ho2020_diffusion}, with the Adamax optimizer and a cosine learning rate schedule with a peak learning rate of $10^{-4}$ and and 500 warmup steps.

\subsection{Software}

This research made use of CAMB 1.3.6 \citep{lewis2002_camb}, PyTorch \citep{pytorch}, JAX \citep{jax}, NumPy \citep{numpy}, Matplotlib \citep{matplotlib}, TensorBoard \citep{abadi2016_tensorflow}, and Weights\&Biases \citep{wandb}. We acknowledge the use of ChatGPT for guidance and clarification during the course of this research.

\clearpage


\end{document}